\documentclass[aps,prb,twocolumn,amssymb,showpacs,superscriptaddress]{revtex4}
\usepackage{graphicx}
\begin{document}
\title{Spin polarization of light atoms in jellium: Detailed electronic structures}
\author{V.~U.~Nazarov}
\affiliation {Department of Physics and Institute for Condensed Matter Theory,
Chonnam National University, Gwangju 500-757, Korea}
\author{C.~S.~Kim}
\affiliation {Department of Physics and Institute for Condensed Matter Theory,
Chonnam National University, Gwangju 500-757, Korea}
\author{Y.~Takada}
\affiliation {Institute for Solid State Physics, University of Tokyo,
Kashiwa, Chiba 277-8581, Japan}
\date\today
\begin{abstract}
We revisit the problem of the spontaneous magnetization  of an
{\em sp} impurity atom in a simple metal host. The main features of
interest are: (i) Formation of the spherical spin density/charge
density wave around the impurity; (ii) Considerable decrease in the
size of the pseudoatom in the spin-polarized state as compared with
the paramagnetic one, and (iii) Relevance of the electron affinity
of the isolated atom to this spin polarization, which is clarified
by tracing the transformation of the pseudoatom into an isolated
negative ion in the low-density limit of the enveloping electron
gas.
\end{abstract}

\maketitle

%%%%%%%%%%%%%%%%%%%% Introduction %%%%%%%%%%%%%%%%%%%%%%%%%%
Interests in spintronics are on the rise from both scientific and technological
points of view. \cite{Wolf-02,Zutic-04} Since devices in spintronics involve
active control and manipulation of spin degrees of freedom in solid-state
systems, it is absolutely necessary to have a deeper understanding of fundamental
interactions between electron spins and its solid-state environments. In view of
this situation, we are interested in a composite system of an atom immersed into
the otherwise homogeneous electron gas (EG).

In an isolated atom, the ground state obeys the Hund's multiplicity rule that
requires the highest spin configuration compatible with the Pauli's exclusion
principle. Physically this rule is interpreted as the
consequence of an effectively larger nuclear charge in a higher spin
configuration due essentially to the exchange effect. \cite{Boyd-84}
Similarly in a uniform EG, the same effect favors spin polarization, bringing
about the spontaneous spin-symmetry breaking or the spin-density-wave state
which was proven to be the ground state at arbitrary electron densities within
the Hartree-Fock (exchange only) approximation. \cite{Overhauser-60,Overhauser-62}
The correlation effect, however, acts in the opposite direction \cite{Overhauser-62}
and this effect is so strong in an EG as to lead eventually to the paramagnetic
ground state for the majority of metals.

This paper deals with the composite system of an atom immersed into EG.
Investigation of atoms embedded in the EG in both their paramagnetic
\cite{Almbladh-76, Norskov-80,Stott-80,Puska-81,Puska-91,Menchini-03}
and spin-polarized  \cite{Nieminen-80,Stefanou-91,Papanikolaou-92,Papanikolaou-93,Mavropoulos-98}
states has a long history. However, to the best of our knowledge, some important
features of the electronic structure of the spontaneously
spin-polarized states of this system have not been addressed so far.
More specifically, they include: (i) Formation of the spherical combined
spin density/charge density wave, which slowly decays with the distance
from the impurity; (ii) Significant shrinkage of spin-polarized pseudoatoms
as compared with their spin-neutral counterparts, and (iii) Demonstration of the way
how the spin-polarized states of the impurities turn into those of the negative
ions of the corresponding isolated atoms as the density of the enveloping EG
tends to zero. The purpose of this work is to elucidate the above points.

%%%%%%%%%%%%%%%%%%%% Formulation %%%%%%%%%%%%%%%%%%%%%%%%%%
We are concerned with  an impurity of the atomic number $Z$ (a pseudoatom)
embedded into the otherwise homogeneous EG at zero temperature characterized
by its electron-density parameter $r_s=(3/4\pi n_0)^{1/3}$, where $n_0$ is the
uniform density of the EG in the absence of the impurity. In the spin-density
functional theory (SDFT), \cite{Gunnarsson-76} the Kohn-Sham equation is
written in atomic units as
\begin{eqnarray}
\left[- (1/2) \, \Delta + v^{eff}_\sigma(r)\right] \psi_{i,\sigma}({\bf r})
= \epsilon_i \psi_{i,\sigma}({\bf r}),
\label{Schr}
\end{eqnarray}
where the spin index $\sigma$ takes either $\uparrow$ or $\downarrow$,
$\epsilon_i$ and $\psi_{i,\sigma}$ are, respectively, the energy level and
the wave function of a Kohn-Sham electron orbital, $v^{eff}_\sigma(r)$
given by
\begin{eqnarray}
v^{eff}_\sigma(r)&=& - Z/r + \int  [n(r')-n_0]/|{\bf r}-{\bf r'}| \, d{\bf r}'
\label{veff}\\
&+&v^{xc}_\sigma([n_\uparrow,n_\downarrow];r)-v^{xc}(n_0)
\nonumber
\end{eqnarray}
is the effective potential, where $n(r)=n_\uparrow(r)+n_\downarrow(r)$ is
the local electron density, $v^{xc}_\sigma([n_\uparrow,n_\downarrow];r)$
defined as
\begin{eqnarray}
v^{xc}_\sigma([n_\uparrow,n_\downarrow];r)=
\delta E^{xc}[n_\uparrow,n_\downarrow]/\delta n_\sigma(r)
\end{eqnarray}
is the spin-dependent exchange and correlation (xc) potential with
$E^{xc}[n_\uparrow,n_\downarrow]$ being the total xc energy of the system, and
$v^{xc}(n_0)$ is the spin-independent xc potential at the uniform electron
density $n_0$. The spin densities are  self-consistently determined as
\begin{equation}
n_\sigma(r)= \sum\limits_i |\psi_{i,\sigma}({\bf r})|^2.
\label{n}
\end{equation}

The energy  of a pseudoatom is the difference between the energies of the EG
with and without the impurity:
\begin{widetext}
\begin{eqnarray}
E&=&\sum\limits_{i \in bs} \epsilon_i
+ (1/ 2 \pi) \sum\limits_{l,\sigma} (2 l +1)
\int\limits_0^{k_f} k^2 \, \delta'_{l,\sigma}(k) d k +  \int \left\{Z [n_0-n(r)]/r
- \sum\limits_\sigma v^{eff}_\sigma(r) n_\sigma(r)\right\} d{\bf r}
\cr\cr
&+& (1/2) \int  [n(r)-n_0] [n(r')-n_0]/|{\bf r}-{\bf r'}| \, d{\bf r} d{\bf r}'
+ \int \left\{n(r) \epsilon^{xc}([n_\uparrow,n_\downarrow];r)- n_0
\epsilon^{xc}(n_0) \right\} d{\bf r},
\label{E}
\end{eqnarray}
\end{widetext}
where $\delta'_{l,\sigma}(k)$ stands for the derivative of the phase-shift
of the angular momentum $l$ of the wave-function for a state in the continuous
spectrum in the potential in Eq.~(\ref{veff}). In Eq.~(\ref{E}) the first term
represents the contribution from the bound states, the second term comes from
the change in the density of continuum states due to the interaction with the
impurity, while all the rest are ordinary (S)DFT contributions to the total
energy \cite{Kohn-65} regrouped to insure the convergence of integrals.

%%%%%%%%%%%%%%%%%%%% Calculated Results %%%%%%%%%%%%%%%%%%%%%%%%%%
\begin{figure}[h]
\includegraphics[width=0.45\textwidth,height=0.3375\textwidth]{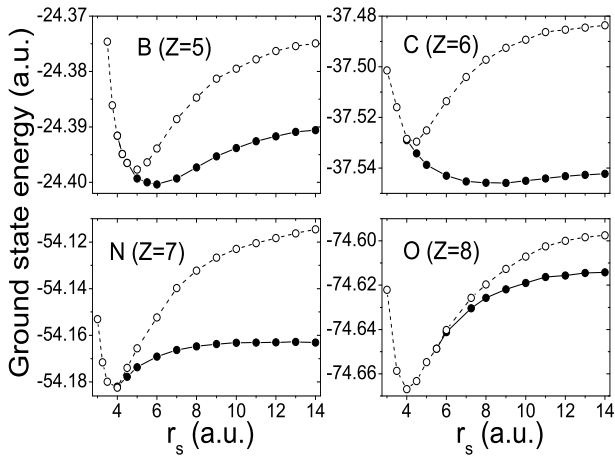}
\caption{Total energy of the spin-polarized (solid curves) and spin-neutral
(dashed curves) states of the B, C, N, and O pseudoatoms versus the EG density
parameter $r_s$.}
\label{enf}
\end{figure}

We have solved Eqs.~(\ref{Schr})-(\ref{n}) self-consistently for the atoms
in the first two rows of the periodic table immersed into the EG of various
densities. For H, He, Li, Be, F, and Ne pseudoatoms, spin-neutral ground
states have been found in the EG density range of $3\le r_s \le 14$.
For B, C, N, and O pseudoatoms, on the other hand, we found spin-polarized
ground states at the density of the EG lower than a certain threshold values,
while the ground state was spin-neutral at higher EG densities. These conclusions
agree with those of earlier studies. \cite{Papanikolaou-93}

In Fig.~\ref{enf} we plot the total energy of Eq.~(\ref{E}) of the spin-polarized
and spin-neutral lowest-energy states of the B, C, N, and O pseudoatoms within
the local spin-density approximation (LSDA) to the SDFT using the parametrization
of the correlation energy of Ref.~\onlinecite{Perdew-81}. In all the four cases,
below a definite threshold value of the EG density, which is different for
different impurity atoms, the spin-polarized ground state has persistently
lower total energy compared with its unpolarized counterpart.

Our method of breaking the spin-symmetry was to start with imposing the occupancy
of the $2p$ bound state with 3 electrons with spin up and less than 3 electrons
with spin down. Then we let the system relax self-consistently to its
ground-state. No unoccupied bound states would remain upon the achievement
of self-consistency: The $2p$ bound states we had had partially filled would
disappear in the self-consistent potential for spin-down electrons.
For spin-up electrons, depending upon the sort of the impurity atom and the
density of EG, this state would either remain and then be filled with 3
electrons, or it would disappear as well. The net spin polarization would
remain finite in either case.

\begin{figure}[h]
\includegraphics[width=0.45\textwidth,height=0.3375\textwidth]{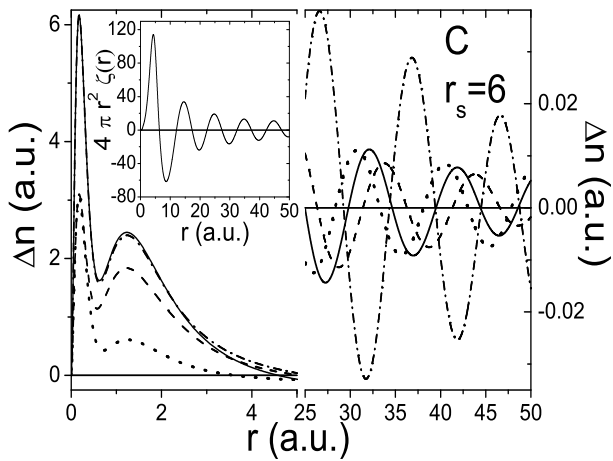}
\caption{Deviation of the density of electrons with spin up, spin down, and
the total electron density from $n_0/2$, $n_0/2$, and $n_0$ (dashed, dotted,
and solid curves), respectively, around the C atom in EG of $r_s=6$.
The dashed-dotted curve represents the unpolarized calculation. The inset
shows the local polarization of Eq.~(\ref{zil}). All curves are multiplied
by $4\pi r^2$.
\label{denf}}
\end{figure}

The results of the calculated spin densities for the carbon atom in the EG of
$r_s=6$ are shown in Fig.~\ref{denf}, together with the total electron density
of the polarized as well as the unpolarized system. We note that at larger
distances from the center, the amplitude of the Friedel oscillations of the
total density in the spin-polarized state is significantly smaller than that
in the neutral state, resulting in the effectively more compact pseudoatom.
The latter finding is consistent with results for isolated atoms. \cite{Boyd-74}
The inset in Fig.~\ref{denf} shows the local polarization
\begin{eqnarray}
\zeta( r)=[n_\uparrow( r) - n_\downarrow(r)]/[n_\uparrow(r) + n_\downarrow(r)].
\label{zil}
\end{eqnarray}

The oscillating and slowly decaying local spin polarization around the impurity
together with Friedel oscillations of the charge-density represent a spherical
combined charge-density/spin-density wave. We determine the total electronic
spin of the pseudoatom as
\begin{equation}
S= (1/2) \, \int [n_\uparrow(r) - n_\downarrow(r)] \, d {\bf r}.
\label{s}
\end{equation}

\begin{figure}
\includegraphics[width=0.45\textwidth,height=0.3375\textwidth]{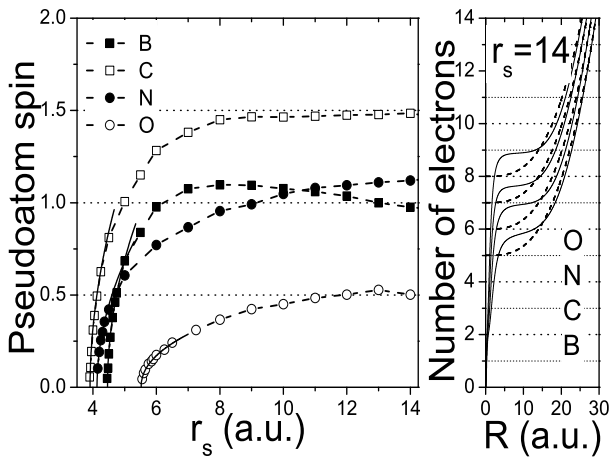}
\caption{Left: Spin of an impurity versus the EG density parameter $r_s$.
Solid lines are the fittings of the data with Eq.~(\ref{power}).
Right: The number of electrons in the sphere of radius $R$ for pseudoatoms
in EG (solid lines); $Z+(R/r_s)^3$ (dashed lines).}
\label{sf}
\end{figure}

In Fig.~\ref{sf} (left panel), the total spin of Eq.~(\ref{s}) is plotted
against the electron-density parameter $r_s$. We conclude that there exists
a finite net spin excess or spontaneous magnetization of the impurity in
the EG at electron densities below the threshold values. The net electronic
{\em spin} of Eq.~(\ref{s}) depends on both the atomic number of the impurity
atom and the EG density, which finds itself in contrast with the result for
the net {\em charge} of the impurity: Due to the full screening of a charge
in the EG, which is closely related to the Friedel sum rule, the pseudoatom
charge is
\begin{equation}
-Z= -\int [n_\uparrow(r) + n_\downarrow(r)-n_0] \, d {\bf r},
\label{Z}
\end{equation}
which is uniquely determined by the sort of the impurity.

While at intermediate densities of the EG the total spin of a pseudoatom
is governed by complicated many-body interactions within the impurity
atom-EG system, the trend in a pseudoatom's spin at low densities (large
$r_s$) has a clear qualitative interpretation. Because of the positive
electron affinity (EA) of the B, C, and O isolated atoms ($0.010$, $0.046$,
and $0.054$ a.u., respectively, \cite{Hotop-85}) the limiting case of
these atoms immersed into the EG at zero EG density are the negative ions (NI)
of the corresponding atoms. According to the Hund's rule, the populations
of the $2p$ orbital are with 2 electrons with spin up ($^3$P), 3 electrons
with spin up ($^4$S), and 3 electrons with spin up and 2 electrons with
spin down ($^2$P) for B$^-$, C$^-$, and O$^-$ ions, respectively,
corresponding to the total spin of 1, 3/2, and 1/2, respectively, which is
clearly satisfied in Fig.~\ref{sf} at large $r_s$. On the other hand, the
NI of the N atom is unstable although long living (EA=$-0.003$ a.u.
\cite{Hotop-85}), and the slow growth of the spin of this pseudoatom between
1 and 3/2 at large $r_s$ can be understood as the competition between the
NI $^3$P and atomic $^4$S states.

In the right panel in Fig.~\ref{sf}, the integrated number of electrons
in a sphere of radius $R$ are plotted versus the radius of the sphere for
the EG of $r_s$=14. The plateaus in the case of B, C, and O close to the
number of electrons of 6, 7, and 9, respectively, prove unambiguously the
NI character of the corresponding states, while for N this
number is between 7 and 8, inferring a state intermediate between an atom
and NI. The growth in the number of electrons to the right from
plateaus is due to the electron density approaching the constant value
of $n_0$ at large distances from the center. This figure also shows that
for a low-density EG, electrons extra to an atom or NI,
whichever supported in the zero-density limit, are pushed away from
the center leaving a region of nearly zero electron density between
the atom/ion and the region of nearly uniform EG, where the number of
electrons in the sphere of radius  $R$ is approximately $Z+(R/r_s)^3$
(dashed curves).

For the period 1 and the rest of the period 2 atoms the same arguments lead
to the spin-neutrality of the corresponding pseudoatoms: In the case of H,
Li and F, which also have positive EA, the acquisition of an extra electron
completes the outer shells, causing the corresponding pseudoatoms to be
spin-neutral in the low-density limit. For a different reason but to the
same effect, in the case of He, Be and Ne, the spin-neutrality holds because
of the non-existence of their NI (even unstable ones with
sufficiently long life-time \cite{Hotop-85}), while the corresponding
atomic states have zero spin.

The steep fall in the spin of a pseudoatom near the critical point seen
in Fig.~\ref{sf} is suggestive of a phase transition of the second order
with the power dependence of the spin on $r_s$ near its critical value $r_{sc}$
\begin{equation}
S\approx a (r_s-r_{sc})^\mu.
\label{power}
\end{equation}
\noindent
In Table~\ref{tab}, the best fit values of the parameters in Eq.~(\ref{power})
are listed. These value strongly suggest that the exponent $\mu$ is universal
and equal to 0.5. \cite{Stefanou-91}
\begin{table}[h]
  \centering
\begin{tabular}{c c c c c}
  \hline
  Atom & B & C & N & O \\
  \hline
  $r_{sc}$ & 4.46 & 3.91  & 4.13 & 5.52\\
  $\mu$    & 0.50 & 0.52  & 0.50 & 0.46\\
  $a$      & 0.94 & 1.11  & 0.73 & 0.24\\
  \hline
\end{tabular}
  \caption{ Best fit parameters in Eq.~(\ref{power}).}
  \label{tab}
\end{table}

At the intermediate EG densities between the threshold value and zero,
the total spin of a pseudoatom obtained via Eq.~(\ref{s}) is not, generally
speaking, a multiple of 1/2, as seen in Fig.~\ref{sf}. This fundamental
difference between an isolated atom and the present pseudoatom is brought
about by the contribution of the infinite number of delocalized electrons
in the latter case. \footnote{Because of the delocalized electrons,
we can follow the convergence of a pseudoatom to NI
within  LDA, while for an isolated NI  LDA fails:
\cite{Perdew-81} The screening at large distances eliminates the
difficulty of LDA for NI, i.e., an electron feeling the -1 charge
if the self-interaction is not subtracted.}

\begin{figure}
\includegraphics[width=0.44\textwidth,height=0.32\textwidth]{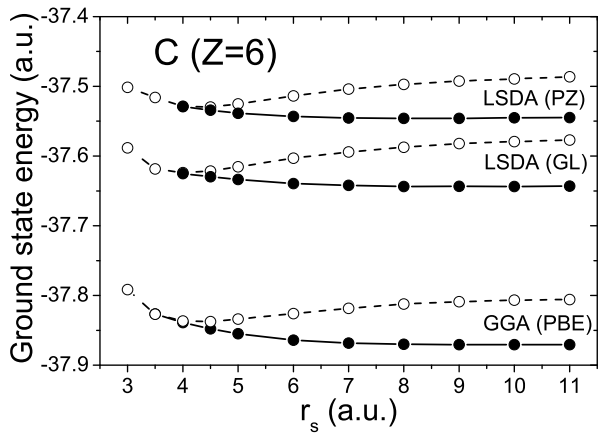}
\caption{
Sensitivity of the energy of the C impurity to the choice of the xc
potential. Solid (dashed) curves refer to the polarized (neutral) states.
The xc potentials used: LSDA of Ref. \onlinecite{Perdew-81} (PZ) and
of Ref. \onlinecite{Gunnarsson-76} (GL), and GGA of Ref. \onlinecite{Perdew-96} (PBE).}
\label{GGA}
\end{figure}

In order to make connection to the earlier works as well as to test the
sensitivity of our results to the  choice of the xc potential, we have
repeated the calculations for the C pseudoatom within LSDA using the
parametrization of the xc energy of Ref. \onlinecite{Gunnarsson-76} and
also beyond LSDA within the generalized gradient approximation (GGA) in
its PBE version. \cite{Perdew-96} In the former case our results for the
unpolarized states reproduce those of Ref.~\onlinecite{Puska-81}. As
shown in Fig.~\ref{GGA}, regardless of the choice of the xc potential,
we have been able to obtain the spin polarized ground state of an atom
embedded in the EG. While the total energies of the both polarized and
unpolarized states are shifted depending on a specific approximation,
their difference (i.e., the stabilization energy of the polarized state)
does not show considerable sensitivity to the choice of the xc potential.

In conclusion, we have performed the spin-density functional calculation
of the spin states of the period 1 and 2 atoms embedded in electron gas.
For H, He, Li, Be, F, and Ne pseudoatoms, we have obtained the spin-neutral
ground states in a wide density range of the electron gas. On the contrary,
for B, C, N, and O pseudoatoms, there occurs a transition into the
spin-polarized state at a critical density of the electron gas which
depends on the atomic number of the impurity. Both results are in accord
with earlier studies. In the spin-polarized state, the pseudoatom is found
to be of a smaller effective size compared with its spin-neutral counterpart,
which is a feature in common with isolated atoms. We also observe a combined
spherical spin-density/charge-density wave which manifests itself as
the Friedel-like oscillations. In the limit of the low density of
the electron gas the electronic structure of a pseudoatom is found to
converge to that of the negative ion of the corresponding isolated atom.
The electronic structure of the spin-polarized state is largely different
from that of the spin-neural one, which will certainly have impact on such
applications as the stopping power of metals for ions and the residual
resistivity of alloys.

VUN and CSK acknowledge support by the Korea Research
Foundation by Grant No. KRF-2003-015-C00214.

\end{document}